\documentclass[pra, superscriptaddress, floatfix, twocolumn]{revtex4-2}
\usepackage[utf8]{inputenc}
\usepackage[T1]{fontenc}

\usepackage{amsfonts,amscd,amsmath,amsthm}
\usepackage{enumitem}
\usepackage{epsfig}
\usepackage{subfigure}
\usepackage{xcolor}
\usepackage[colorlinks = true]{hyperref}
\usepackage{physics}
\usepackage{epstopdf}
\usepackage{framed}
\usepackage{multirow}
\usepackage{color}
\usepackage{longtable}
\usepackage{comment}
\usepackage[ruled,vlined]{algorithm2e}
\usepackage[most]{tcolorbox}

\graphicspath{ {./img/} }

\usepackage{pifont}
\usepackage{bm}

\newtheorem*{theorem*}{Theorem}

\begin{document}
\title{Distilling the knowledge with quantum neural networks}

\author{Yuxuan Yan}
\thanks{These two authors contributed equally to this work.}
\email{yanyx21@mails.tsinghua.edu.cn}
\affiliation{%
Center for Quantum Information, Institute for Interdisciplinary\\
Information Sciences, Tsinghua University, Beijing, China
}%

\author{Sitian Qian}
\thanks{These two authors contributed equally to this work.}
\email{sitian.qian@cern.ch}
\affiliation{%
Department of Physics, University of Wisconsin-Madison, USA
}%

\author{Qi Zhao}
\email{zhaoqi@cs.hku.hk}
\affiliation{%
QICI Quantum Information and Computation Initiative, School of Computing and Data Science, The University of Hong Kong, Hong Kong SAR, China
}%

\author{Xingjian Zhang}
\email{zxj24@hku.hk}
\affiliation{%
QICI Quantum Information and Computation Initiative, School of Computing and Data Science, The University of Hong Kong, Hong Kong SAR, China
}%

\begin{abstract}

Quantum Neural Networks (QNNs) are a promising class of quantum machine learning models with potential quantum advantages when implemented on scalable, error-corrected quantum computers. However, as system sizes increase, deploying QNNs becomes challenging. Similar to their classical counterparts, a key obstacle to their practical applications is that large-scale QNNs may not be easily deployed on smaller systems that have limited resources. Here, we tackle this challenge by compressing QNNs via knowledge distillation. We demonstrate how well-trained QNNs on large systems can be distilled into smaller architectures with similar configurations. We numerically show that knowledge distillation helps reduce the training cost of QNNs in terms of the number of qubits and circuit depth. Additionally, we find that a self-knowledge-distillation approach can accelerate training convergence. We believe our results offer new strategies for the efficient compression and practical deployment of QNNs.

\end{abstract}

\flushbottom
\maketitle

\thispagestyle{empty}

\section{Introduction}
Machine learning is a transformative approach that enables computing systems to learn from data and make autonomous predictions. Among various machine learning models, neural networks (NNs) stand out by mimicking the structure and functionality of the human brain, thereby providing effective and accurate processing of large datasets \cite{mcculloch1943logical,rosenblatt1958perceptron,werbos1974beyond,lecun1989backpropagation}. Concurrently with the progress in neural network theory, advances in quantum information technologies have highlighted quantum phenomena as promising resources to enhance computational efficiency, exemplified by the acceleration of integer factorization in Shor's quantum algorithm \cite{shor_polynomial-time_1997}. Inspired by the computational power of quantum devices, an interdisciplinary model known as the Quantum Neural Network (QNN) has emerged, aiming to combine the strengths of both quantum computing and neural networks \cite{cerezo_variational_2021}.

Compared to their classical counterparts, QNNs are naturally capable of learning from both quantum and classical information. When processing quantum inputs, QNNs directly receive quantum states, manipulate them through parameterized quantum circuits, and generate predictions about the properties of these input states via measurements. For classical inputs, an additional step is implemented to encode the classical data into quantum states, after which the same processing procedure for quantum inputs can be applied. Owing to their unique information processing capabilities, QNNs have found diverse applications \cite{cerezo_variational_2021}, with particular success in classification tasks \cite{cong_quantum_2019, herrmann_realizing_2022}. QNNs offer significant advantages over their classical counterparts, particularly in processing quantum input data, making them especially promising for applications such as quantum sensing \cite{degen_quantum_2017, raphael_optimal_2023}. Moreover, the potential for QNNs to outperform classical neural networks in handling classical data remains an active area of research. Preliminary theoretical results suggest that QNNs may exhibit superior generalization capabilities compared to classical NNs \cite{abbas_power_2021, caro_generalization_2022}.

The performance of QNNs is strongly dependent on the scale and quality of available quantum computing hardware. Recent experimental advancements \cite{zhao_realization_2022, google_quantum_ai_suppressing_2023, bluvstein_logical_2023, acharya_quantum_2024}, along with quantum computing roadmaps from leading technology companies \cite{quantinuum2024,ibmroadmap2025}, indicate that early fault-tolerant quantum platforms featuring thousands of qubits may become available in the near term. These developments pave the way for constructing large-scale QNNs with substantial potential for practical, real-world applications. Notwithstanding, while we look forward to the full potential of quantum machine learning, a foreseeable challenge lies in deploying large-scale models on limited quantum resources. Drawing from our experiences in classical high-performance computation, it is expected that access to large-scale quantum platforms will initially be restricted to a few organizations. Although QNN models running on such platforms demonstrate remarkable learnability and expressivity, they may be impractical for deployment on resource-constrained devices, potentially exceeding their computational capacities. This challenge underscores the importance of model size reduction and knowledge transfer to users with smaller quantum computers.

\begin{figure*}[hbtp!]
    \centering
    \includegraphics[width=\textwidth]{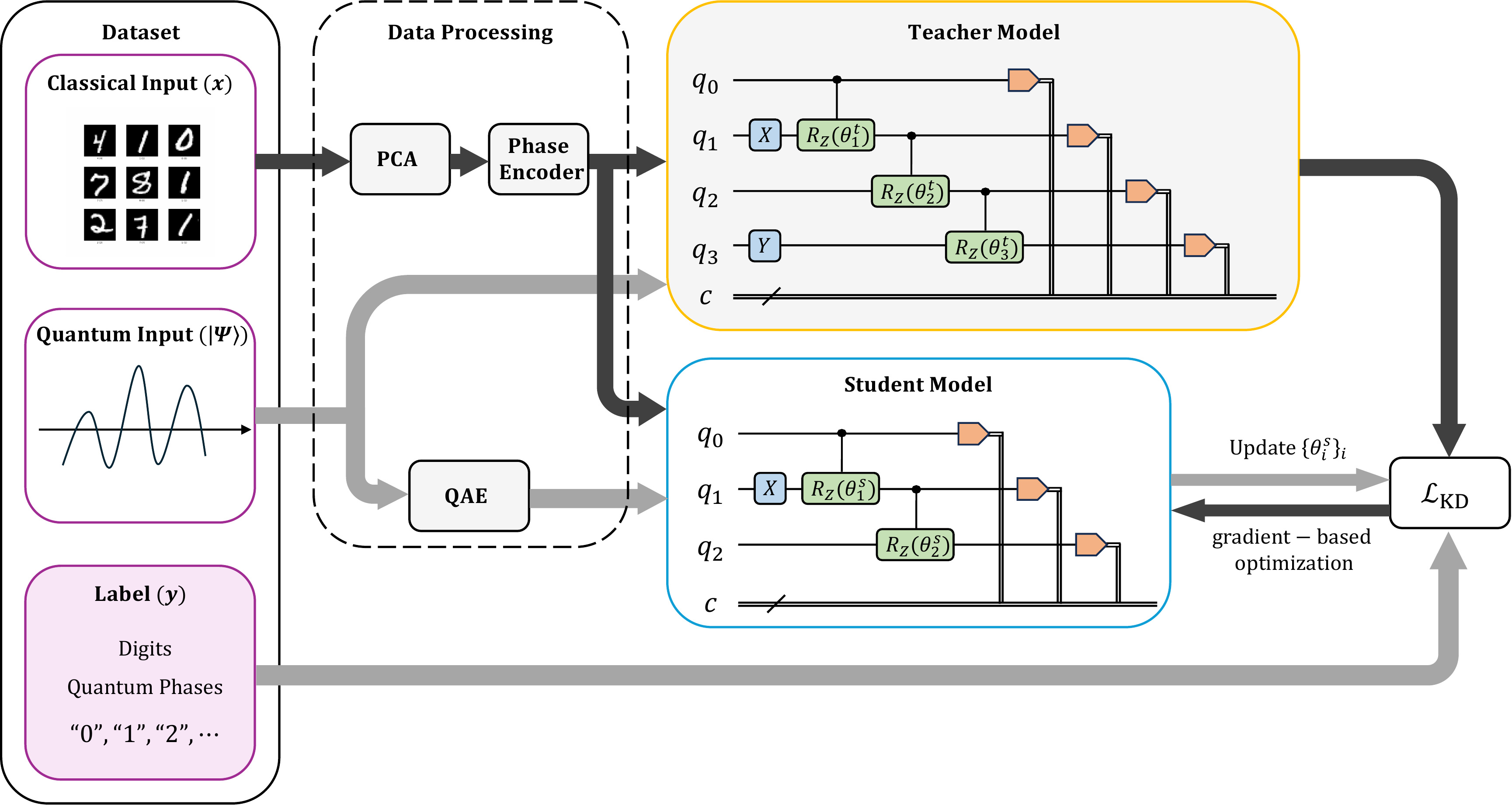}
    \caption{Illustration of the proposed QNN-KD framework. The dataset is processed by both a teacher model and a student model. The dataset comprises input data and label. The input can be either classical data represented as vectors, $x$, or quantum data represented as states, $\ket{\psi}$.
    The labels, $y$, may represent different physical meanings, such as the number that a handwritten digit image represents and various quantum phases.
    Before feeding the input data into the QNN models, data processing may be required. The classical input is first processed with Principal Component Analysis (PCA) and then encoded into quantum states. In our work, we apply a quantum phase encoder. The quantum input may be processed with a Quantum AutoEncoder (QAE) to match the size of the QNN. The teacher model generates supervisory signals. The student QNN model learns to mimic the teacher's outputs by minimizing the knowledge distillation loss $\mathcal{L}_{\mathrm{KD}}$ over a set of tunable parameters $\{\theta_i^s\}_i$ through gradient-based optimization. In this work, we consider two cases: the teacher model is a larger QNN than the student model with a similar architecture, and the teacher model is the same as the student model for self-KD. QNN-KD can efficiently reduce the size of the QNN and accelerate training in a self-KD scenario.}
    \label{fig:intro}
\end{figure*}

In this work, we propose an efficient method for transferring knowledge from large QNNs. Our approach is inspired by Knowledge Distillation (KD), a widely used technique for model compression in classical machine learning \cite{hinton_distilling_2015,gou_knowledge_2021}. Our main results demonstrate that KD can effectively compress QNNs by transferring knowledge from large QNNs acting as teachers to smaller QNNs serving as students, achieving efficient model compression while preserving performance. Additionally, we investigate a variant of this method called Self-Knowledge Distillation (self-KD), where the teacher and student are the same QNN. An illustration of the proposed workflow, referred to as QNN-KD, is presented in Figure~\ref{fig:intro}.

\section{Theory}
\subsection{Preliminaries for quantum neural networks}
\begin{figure*}[hbt!]
    \centering
    \includegraphics[width=0.65\textwidth]{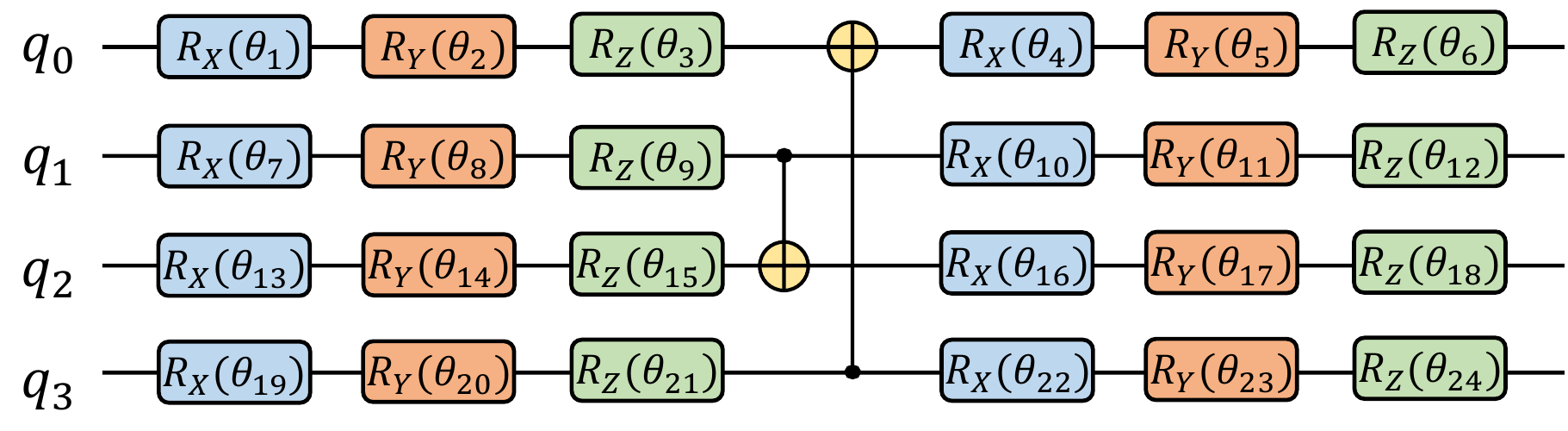}
    \caption{Hardware-efficient QNN architecture illustrated with four qubits and a single layer. The single-qubit rotations consist of sequential rotations around the $X$-, $Y$-, and $Z$-axes, as denoted by blue, orange, and green boxes, respectively. The entangling operations are implemented using controlled-NOT (CNOT) gates. After applying layers of single-qubit rotation gates around each of the three axes, a layer of CNOT gates that entangle multiple qubits are applied. The circuit architecture proceeds with alternate layers between single-qubit gates and CNOT gates. The circuit depth is defined as the number of CNOT gate layers.}
    \label{fig:qnn_block}
\end{figure*}

We first review the basic concepts of the QNN that are necessary for our work. As shown in the box on the left of Figure~\ref{fig:intro}, a QNN may take as inputs the classical data that are represented as vectors, $\{\bm{x}^{(i)}\}_{i=1}^{N}$, or quantum data that are represented as quantum states, $\{\rho^{(i)}\}_{i=1}^{N}$, after which it predicts specific properties $\{y^{(i)}\}_{i=1}^{N}$ correspondingly. The execution of QNNs is divided into three steps:

\begin{enumerate}[leftmargin=*]
    \item For classical inputs, an encoding circuit is applied to map the data into quantum states $\rho^{(i)}$ such that a quantum computer can process.

    \item A parametric circuit $U(\bm{\theta})$ is applied to $\rho^{(i)}$, with a set of parameters $\bm{\theta}$ to be trained. Various parametric circuit architectures can be utilized. In general, the parametric circuits are allowed to comprise any non-parametric gates and parametric rotation gates like single-qubit rotation gates:
    \begin{equation}
        R_X(\theta)=e^{i\theta X},R_Y(\theta)=e^{i\theta Y},R_Z(\theta)=e^{i\theta Z}.
    \end{equation}
    In this work, we use a hardware-efficient architecture as illustrated in Figure~\ref{fig:qnn_block}, which alternates between layers of single-qubit rotation gates and layers of multi-qubit entangling gates.

    \item After applying quantum circuits, we measure the expectation value of a set of observables, $O_j,j \in \{1,2,\cdots,m\}$, and take them as logits $l_j$:
    \begin{equation}
        l_j^{(i)} = \langle O_j^{(i)} \rangle \big| _{\bm{\theta}} = \tr [O_j U(\bm{\theta}) \rho^{(i)} U(\bm{\theta})^\dagger].
        \label{eq:logits}
    \end{equation}
\end{enumerate}

At the end of an $m$-class classification task, the QNN model outputs probabilities $\Pr(y^{(i)}=j \mid \bm{x}^{(i)})$ for $j \in \{1,\dots,m\}$ by applying a softmax function $\sigma_j$ to the logits:
\begin{equation}\label{eq:softmax}
    \hat{p}_j^{(i)} = \sigma_j \left( l_1^{(i)}, \dots, l_m^{(i)} \right),
\end{equation}
where $\sigma_j(z_1, \dots, z_m) = \frac{e^{z_j}}{\sum_{k=1}^m e^{z_k}}$.
QNNs are trained to predict labels by iteratively updating the parameters to minimize a predefined loss function on the training dataset. Consider a training dataset $\{(\bm{x}^{(i)}, y^{(i)})\}_{i=1}^N$ of input-output pairs. For brevity, denote $\bm{X} = \{\bm{x}^{(i)}\}_{i=1}^N$ and $\bm{y} = \{y^{(i)}\}_{i=1}^N$. To measure the discrepancy between the predictions and the true labels, we can apply the cross-entropy loss, which is given by
\begin{equation}
    \mathcal{L}_{\mathrm{CE}}(\bm{X}, \bm{y}) := -\sum_{i=1}^N \log \hat{p}_{y^{(i)}}^{(i)}, \label{eq:crossent}
\end{equation}
where $\hat{p}_{y^{(i)}}^{(i)}$ is the predicted probability for the true class $y^{(i)}$.
The analytical gradients of the loss function can be obtained by quantum computers. Because the parametric gates are rotation gates, the partial derivatives of expectation values over an element of $\theta_{l}$ can be evaluated by shifting $\theta_{l}$ by $\pm\frac{\pi}{2}$ and taking the difference \cite{li_hybrid_2017, mitarai_quantum_2018, schuld_evaluating_2019, banchi_measuring_2021}:
\begin{equation}
\begin{split}
     \frac{\partial l_j}{\partial \theta_l}&= \frac{\partial \langle O_j \rangle}{\partial \theta_l}\bigg| _{\bm{\theta}} = \frac{1}{2} \left(\langle O_j \rangle\big| _{\bm{\theta}_{l+}} - \langle O_j \rangle\big|_{\bm{\theta}_{l-}}\right),
\end{split}
\end{equation}
with
\begin{equation}
    \bm{\theta}_{l\pm} := (\theta_1, \cdots,\theta_{l-1}, \theta_l\pm\frac{\pi}{2},\theta_{l+1}, \cdots, \theta_{L}),
\end{equation}
where $L$ is the number of parameters in the QNN. Using chain rules, gradients of Eq.~\eqref{eq:crossent} can be obtained, and gradient-based optimization methods can be employed.

\subsection{Knowledge distillation} \label{sec:kd}
Next, we introduce the knowledge distillation (KD) process, which provides a powerful approach to model compression. Suppose one has a well-trained NN model on a large system, which we call the teacher model. For a machine learning task on a smaller yet ``similar'' model (we shall rigorously define the notion of similarities in later sections), one can take it as a student model and transfer the knowledge from the teacher model to it via KD---a method originally proposed by \citet{hinton_distilling_2015}. For this purpose, the distillation loss, $\mathcal{L}_{\mathrm{KD}}$, is defined as:
\begin{equation}\label{eq:kdloss}
\begin{split}
 \mathcal{L}_{\mathrm{KD}}(\bm{X}, \bm{y}; T, \alpha)
 =& \alpha T^2 D_{\mathrm{KL}} \left( \bm{\sigma}\big(l_{\mathrm{s}}(\bm{X})/T\big) \,\big\|\, \bm{\sigma}\big(l_{\mathrm{t}}(\bm{X})/T\big) \right) \\
 &+ (1-\alpha) \mathcal{L}_{\mathrm{CE}} \big(\bm{\sigma}\big(l_{\mathrm{s}}(\bm{X})\big), \bm{y}\big),
\end{split}
\end{equation}
where the loss is summed over samples in $\{(\bm{x}^{(i)}, y^{(i)})\}_{i=1}^N$, and notations are:
\begin{itemize}
 \item $\bm{X} = \{\bm{x}^{(i)}\}_{i=1}^N$ and $\bm{y} = \{y^{(i)}\}_{i=1}^N$ are the input features and labels, respectively;
 \item $D_{\mathrm{KL}}$ is the Kullback-Leibler (KL) divergence between two distributions, where for distributions $P=\{p(x)\}$ and $Q=\{q(x)\}_x$, $D_{\mathrm{KL}}(P\|Q)=\sum_xp(x)\log[p(x)/q(x)]$;
 \item $l_{\mathrm{t}}(\bm{X})$ and $l_{\mathrm{s}}(\bm{X})$ are the logits of the teacher and student models;
 \item $\bm{\sigma}(\cdot)$ is the softmax function introduced in Eq.~\eqref{eq:softmax};
 \item $\alpha \in [0,1]$ balances the two loss terms, and $T > 0$ is the temperature parameter.
\end{itemize}

To be more specific, the distillation loss in Eq.~\eqref{eq:kdloss} consists of two parts: (1) The first part represents the difference between the teacher and student models. The loss is given by the KL divergence, which measures the difference between the predicted probability from the student model, $\bm{\sigma}[l_{\mathrm{s}}(\bm{x})/T]$, and the teacher model, $\bm{\sigma}[l_{\mathrm{t}}(\bm{x})/T]$. Both logits are modulated by the temperature parameter $T$ before being fed into the softmax function. (2) The second part is the classification loss, defined by the cross entropy between the true label, $y$, and the small model logit, $l_\textrm{s}(\bm{x})$.
To achieve the best distillation performance, we shall balance the knowledge from the teacher model with that of the training dataset. Therefore, we introduce a tunable parameter $\alpha$ to adjust this ratio between the aforementioned two parts.

\subsection{Quantum data compression}\label{sec:data-compression}
\begin{figure*}[hbtp!]
    \centering
    \includegraphics[width=\textwidth]{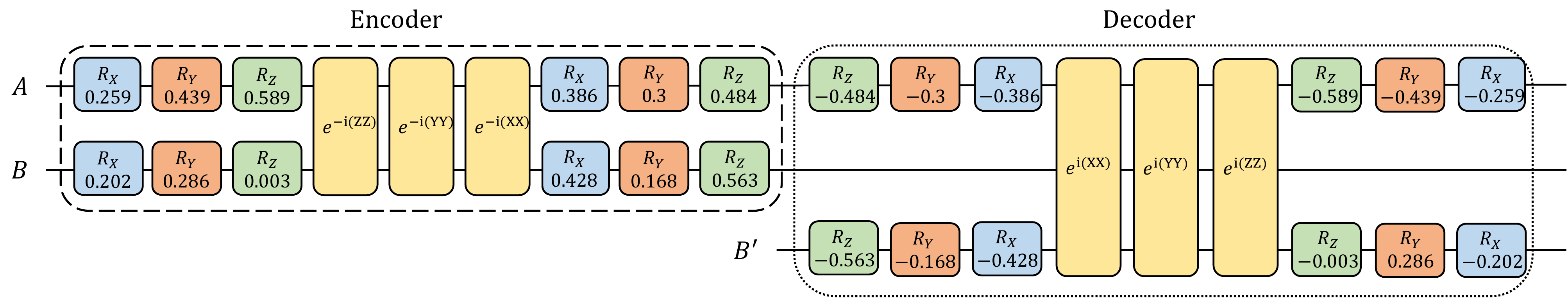}
    \caption{Architecture of the Quantum AutoEncoder (QAE), illustrated with a compression from two qubits to one qubit. In our simulation, we employ a similar architecture for compression. The initial state lies in the joint system given by $A$ and $B$. The state is first encoded into a single-qubit state on system $A$, where we depict the encoder circuit by the dashed box. Afterward, the system is decoded with an auxiliary system, $B'$, where we depict the decoder circuit by the dotted box. The single-qubit rotation gates are defined as in Figure~\ref{fig:qnn_block}, with the values representing those in our numerical experiment. The yellow boxes represent two-qubit entangling gates. The gate $e^{-i(ZZ)}$ represents $\exp[-i(Z\otimes Z)]$ with $Z$ being the Pauli-$Z$ operator, and the other two-qubit gates are defined similarly.}
    \label{fig:qae_arch}
\end{figure*}

In the case of quantum inputs, we need to combine KD with some quantum data compression techniques to reduce the number of qubits of the student model. Here, we employ the Quantum AutoEncoder (QAE) to compress quantum inputs \cite{mangini_quantum_2022}. In essence, the QAE compresses the quantum input data defined on a larger Hilbert space into a lower-dimensional representation suitable for the student model. We depict an example QAE circuit in Figure~\ref{fig:qae_arch}.

Generally, the QAE encodes an input state $\rho^{(i)}$ on a bipartite system, $AB$, into an output state on its first subsystem, $A$. This step is done by first applying a parametric circuit on $AB$ and then tracing out the subsystem $B$:
\begin{equation}
    \rho^{(i)}_{\mathrm{comp}} = \tr_B [U(\bm{\theta}) \rho^{(i)} U(\bm{\theta})^\dagger].
\end{equation}
In practice, tracing out $B$ means measuring the state of subsystem $B$ on a complete and orthogonal basis and summing up the post-measurement state on subsystem $A$ weighted by the measurement outcome probabilities. To decode the compressed quantum data, we need to introduce an auxiliary system, $B^\prime$, whose dimension is the same as $B$, and apply the inverse unitary $U(\bm{\theta})_{AB^\prime}^\dagger$ on system $AB^\prime$:
\begin{equation}
    \rho^{(i)\prime} =  U(\bm{\theta})_{AB^\prime}^\dagger [\rho^{(i)}_{\mathrm{comp}} \otimes \ketbra{0}_{B^\prime}] U(\bm{\theta})_{AB^\prime}.
\end{equation}
To minimize the information loss caused by compression, we need to maximize the average fidelity between the states processed by the QAE, $\rho^{\prime(i)}$, and the original states over the training set, $\rho^{(i)}$:
\begin{equation}
    \bar F(\bm{\theta}) = \frac{1}{N} \sum_{i=1}^{N} F\left(\rho^{(i)}, \rho^{\prime(i)}\right), \label{eq:avfid}
\end{equation}
where the fidelity is given by $F(\rho_1, \rho_2) := \tr \left( {\sqrt {{\sqrt {\rho_1 }}\rho_2 {\sqrt {\rho_1 }}}} \right)^{2}$. By definition, the fidelity is between $0$ and $1$. A fidelity close to one indicates a small information loss during encoding. In the special case where $\rho_1$ is a pure state, i.e., $\rho_1 = \ketbra{\psi}$, the fidelity can be calculated by $F(\rho_1, \rho_2) = \bra{\psi} \rho_2 \ket{\psi}$.

\section{Numerical experiments}
Now, we combine the QNN with KD and present the results of our quantum neural network knowledge distillation (QNN-KD) scheme. In the numerical experiments, we applied the MindSpore Quantum platform \cite{xu2024mindspore} and TensorCircuit \cite{zhang2023tensorcircuit}. We numerically demonstrate KD for QNNs across two distinct scenarios: one with quantum inputs and the other with classical inputs. In the quantum-input scenario, we focus on the classification of quantum topological phases, a central problem in quantum many-body physics~\cite{wen_topological_2013}. In the classical-input scenario, we address the classification of handwritten digit images. Together, these tasks exemplify both the foundational and practical applications of QNNs, demonstrating their versatility across diverse domains. An overview of the QNN-KD framework is shown in Figure~\ref{fig:intro}, and we illustrate its workflow through the following tasks.

In all the experiments, the parameters of QNNs, $\{\theta_l\}_{l=1}^{L}$, are randomly initialized from a uniform distribution on $[-0.1 \pi, 0.1 \pi]$. We employed the Adam optimization algorithm to train the student model parameters. The initial learning rate was set to $0.01$, with $\beta_1=0.9$ and $\beta_2=0.999$, and L2 regularization (weight decay) was set to $0$. Additionally, we used a step learning rate scheduler, which decayed the learning rate by a multiplicative factor of $\gamma=0.5$ every $100$ training steps.

\subsection{Quantum inputs: classification of quantum topological phases}\label{sec:quantumclassification}
We first consider the quantum-input case, where QNNs are used to classify quantum topological phases. The quantum input data are the ground states of a one-dimensional (1D) ZXZ Hamiltonian, defined as
\begin{equation}
    H = - \sum_{i=1}^{n} h_1 X_i + h_2 X_i X_{i+1} + J Z_{i-1}X_{i}Z_{i+1},
\end{equation}
where $n$ denotes the number of qubits in the chain, and $h_1$, $h_2$, and $J$ are real coefficients. Phase classification of this model has been previously demonstrated using quantum convolutional neural networks~\cite{cong_quantum_2019}. When $h_1 = 0$, a phase transition occurs at $\left|\frac{h_2}{J}\right| = 1$. Specifically, for $\left|\frac{h_2}{J}\right| > 1$, the ground state corresponds to a paramagnetic phase (labeled ``0''), while for $\left|\frac{h_2}{J}\right| < 1$, it corresponds to a symmetry-protected topological phase (labeled ``1'').

In our experiments, we construct the dataset by numerically solving for the ground states of random Hamiltonians with $n = 7$ and $n=15$ qubits, fixed coefficients $h_1 = 0$ and $J = 1$, and $h_2$ sampled uniformly from the interval $[0.8, 1.2]$.

In the 7-qubit case, we explore two directions of model compression: (1) the student QNN has the same circuit depth but fewer qubits than the teacher, and (2) the student QNN has the same number of qubits but a shallower circuit depth. For the first direction, where the student model has fewer qubits, we apply the QAE to reduce the dimensionality of the teacher’s quantum input states before passing them to the student. In this experiment, QAE's training takes $600$ iterations. The averaged fidelity of QAE is $0.897$.

The teacher model is a 7-qubit and 2-layer QNN, while the student models have compressed size as listed in Table~\ref{tab:qnn_quantum}. All training runs consist of 50 iterations, using a random batch of 64 data samples for each iteration. The KD hyperparameters are set with a temperature of $T=2$ and a weighting factor of $\alpha=0.8$. As summarized in Table~\ref{tab:qnn_quantum}, our results show that KD enables substantial model compression while preserving high performance in both compression directions. Compared with student models trained without KD, the test accuracy increases from $52.3\%$ to $81.7\%$ in the first scenario and from $86.0\%$ to $99.8\%$ in the second.

\begin{table}[htbp]
    \centering
    \begin{tabular}{c|c|c|c|c}
    \hline
    \hline
      QNN Model Type & Qubits & Layers & Test Accuracy (\%) & KD \\ \hline
        Teacher & 7 & 2 & 98.3 & -- \\
        \hline
        Student (1) & 2 & 2 & 52.3 & $\times$  \\ \hline
        Student (1) & 2 & 2 & 81.7 & $\checkmark$  \\ \hline
        Student (2) & 7 & 1 & 86.0 &  $\times$  \\ \hline
        Student (2) & 7 & 1 & 99.8 & $\checkmark$  \\ \hline
        \hline
    \end{tabular}
    \caption{Performance of QNNs with and without knowledge distillation (KD) on quantum phase classification. Besides the student models, we also list the performance of the teacher model as a benchmark.}
    \label{tab:qnn_quantum}
\end{table}

\begin{figure}[bthp!]
    \centering
    \includegraphics[width=\columnwidth]{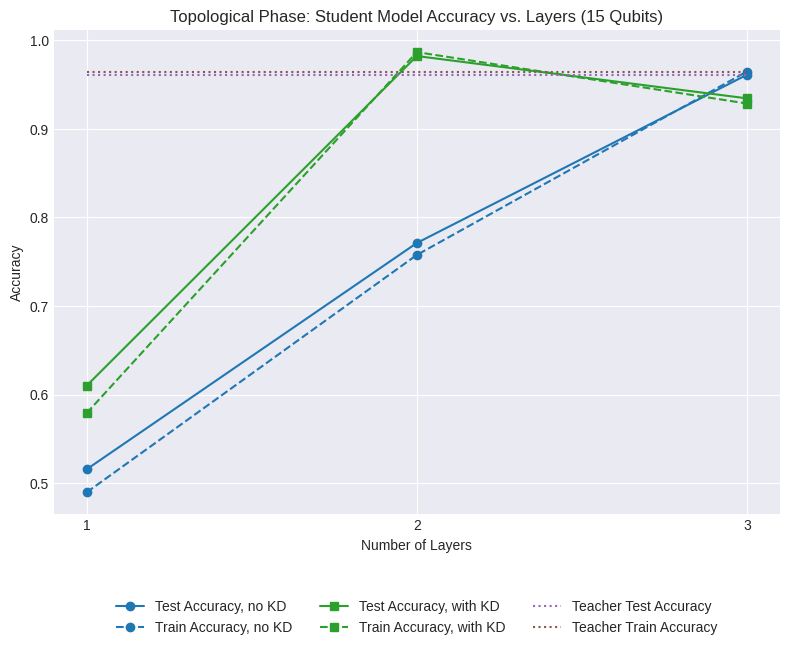}
    \caption{The effect of KD on student model accuracy in topological phase classification as a function of network layers for a 15-qubit system. The plot compares the test (solid lines) and train (dashed lines) accuracies for student models trained without KD (blue) and with KD (green). The constant dotted lines represent the baseline test (purple) and train (brown) accuracies of the larger teacher model.}
    \label{fig:topol}
\end{figure}

We also extend the scope of these experiments to 15 qubits and focus on the reduction of depth. For this experiment, QNNs utilize 15 qubits. The teacher model is a 10-layer QNN, while the student models have layers ranging from 1 to 3. All training runs consist of 50 iterations, using a random batch of 64 data samples for each iteration. The KD hyperparameters are set with a temperature of $T=2$ and a weighting factor of $\alpha=0.8$. Our results are illustrated in Figure~\ref{fig:topol}.

The effectiveness of KD on the topological phase task is highly dependent on the QNN's depth. While KD provides a modest accuracy boost for a single-layer model, its impact is most dramatic at two layers. At this ``sweet spot,'' the KD-trained student achieves near-perfect test accuracy, significantly outperforming both the non-KD student and even the teacher model. This advantage reverses at three layers, where the non-KD model's performance catches up, while the KD model's accuracy slightly drops. This drop is the consequence of overtraining. Our results suggest KD is most beneficial for optimizing intermediate-sized models, but its advantage fades as the student QNN's capacity becomes sufficient on its own.

\subsection{Classical inputs: classification of MNIST images}

The second task is a binary classification using the MNIST dataset \cite{lecun_gradient-based_1998}, which originally contains 28×28 grayscale images of handwritten digits from ``0'' to ``9.'' To construct this binary task, we subset the dataset to include only images of the digits ``1'' and ``5.'' This filtering was applied to the standard, pre-defined MNIST training and test sets. This process yields a final dataset of 12,163 training samples and 2,027 test samples.

To map the high-dimensional classical images to the limited number of available qubits, we employ a data pre-processing pipeline. First, we apply Principal Component Analysis (PCA), a linear dimensionality reduction technique \cite{pearson_liii_1901}, to compress the image data to a lower dimension. Following PCA, the continuous-valued output vector is binarized. This step converts each numerical component into a binary digit by applying a 0.5 threshold: any value greater than 0.5 is set to 1, while any value less than or equal to 0.5 is set to 0. The resulting binary vector is then encoded into a quantum state using a phase encoder, which maps each ``0'' or ``1'' to the phase of a corresponding qubit.

For this experiment, QNNs utilize 15 qubits. The teacher model is a 10-layer QNN, while the student models have layers ranging from 4 to 8. All training runs consist of 350 iterations, using a random batch of 64 data samples for each iteration. The KD hyperparameters are set with a temperature of $T=2$ and a weighting factor of $\alpha=0.8$.

\begin{figure}[bthp!]
    \centering
    \includegraphics[width=\columnwidth]{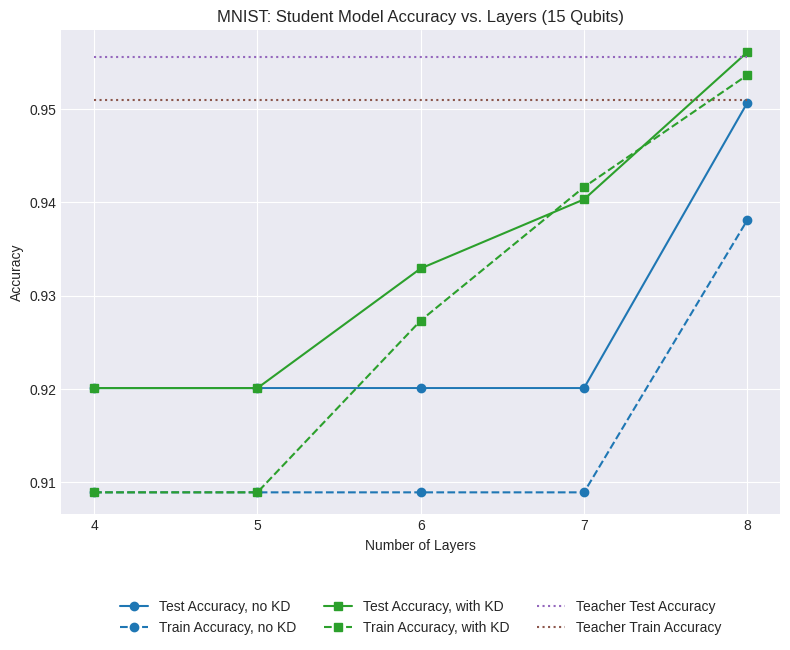}
    \caption{The effect of KD on student model accuracy in MNIST binary classification as a function of network layers for a 15-qubit system. The plot compares the test (solid lines) and train (dashed lines) accuracies for student models trained without KD (blue) and with KD (green). The constant dotted lines represent the baseline test (purple) and train (brown) accuracies of the larger teacher model.}
    \label{fig:mnist}
\end{figure}

We plot our experiment results in Figure~\ref{fig:mnist}. The plot demonstrates that KD provides a significant performance benefit, particularly for student models with moderate depth. The model trained without KD struggles to generalize, showing flat test accuracy at 0.92 until 7 layers, despite its train accuracy remaining low. In contrast, the KD-trained model achieves higher test accuracy across all depths and shows a steady performance increase as layers are added. This suggests KD helps the student model learn more effectively and generalize better, avoiding the performance plateau seen in the non-KD model between 4 and 7 layers. At 8 layers, both models converge to a high test accuracy that rivals the teacher's, indicating that a sufficient model depth is ultimately required to match the teacher's performance.

\subsection{Self-KD for faster convergence}
\begin{figure}[bthp!]
    \centering
    \includegraphics[width=\columnwidth]{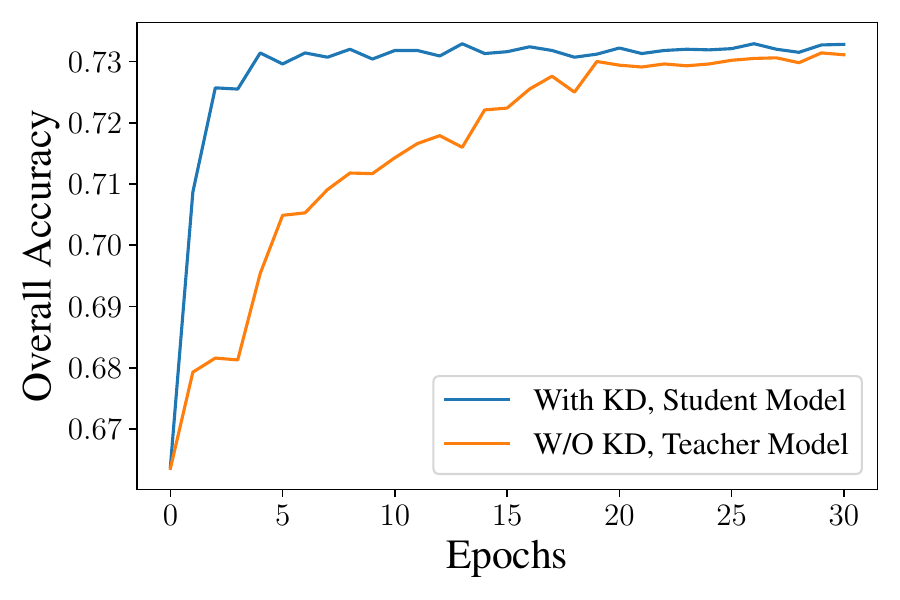}
    \caption{Evolution of the overall accuracy on the test set. With the same architecture and training strategy, the teacher model takes approximately 20 epochs to converge, while the student model takes no more than 5 epochs. }
    \label{fig:multiclass_acc}
\end{figure}

We next investigate the task of Self-Knowledge Distillation (Self-KD) to demonstrate its effectiveness in accelerating the convergence of QNN training. In Self-KD, the teacher and student models share the same architecture, and the knowledge from a pretrained teacher is used to guide the training of a freshly initialized student. In this experiment, we choose the 10-category classification of the MNIST dataset, which was introduced above. We pretrain a 10-qubit, 4-layer QNN as the teacher model, and subsequently train another identically structured QNN as the student under the KD scheme. As shown in Figure~\ref{fig:multiclass_acc}, Self-KD significantly accelerates convergence, as evidenced by the faster improvement of test accuracy during training.

For this task, we employ a state encoder that embeds classical data into the full quantum state vector. The pure state vector of an $n$-qubit system lies in a $2^n$-dimensional Hilbert space. This encoding strategy is adopted for processing QNNs with $28\times28$ full-size MNIST images. Each handwritten digit image is flattened into a 784-dimensional vector and zero-padded to match the 1024-dimensional pure state of a 10-qubit quantum system. Training with and without KD follows the same protocol. The models are trained with a batch size of 64 for 31 epochs. Here, each epoch refers to one complete pass through the entire training dataset. For the KD experiment, the temperature and loss-balance parameters are set to $T=1$ and $\alpha=0.5$, respectively.

\section{Discussion}
In this paper, we generalize KD, a widely recognized model compression technique in classical NNs, to QNNs. Our investigation encompasses QNNs with both quantum and classical inputs, addressing specific application scenarios for each, including the classification of quantum topological phases and the classification of handwritten digits using the MNIST dataset, respectively. Our numerical experiments reveal that, akin to its classical counterpart, KD effectively compresses the model size by facilitating the transfer of knowledge from a teacher model to a student model. Importantly, the teacher model can be a large QNN, a well-trained classical NN with a more complex architecture, or even a QNN with the very same structure for self-KD. In all instances, the student QNN benefits from enhanced training efficiency, manifesting in improved accuracy and reduced time costs.

Currently, our exploration of KD for QNNs is limited to classification tasks. It would be compelling to extend this methodology to other problem types, such as regression and representation learning. Our experiments, in which a classical teacher model assists in training a student QNN with a similar architecture, indicate the potential of a quantum–classical hybrid approach to model compression in quantum machine learning. Since the training cost of classical NNs is substantially lower than that of QNNs, this strategy offers a practical pathway to enhance the scalability and applicability of QNNs.

Furthermore, such a hybrid framework may align well with current noisy intermediate-scale quantum (NISQ)~\cite{preskill_quantum_2018} processors and emerging early fault-tolerant platforms~\cite{zhao_realization_2022, google_quantum_ai_suppressing_2023, bluvstein_logical_2023, acharya_quantum_2024}. Integrating KD with quantum error correction (QEC) protocols could further mitigate noise effects and reduce runtime overheads, improving the robustness of compressed QNNs on realistic hardware. Beyond standalone devices, extending this framework to distributed settings such as the quantum Internet \cite{wehner2018quantum} may enable efficient deployment of lightweight quantum models and facilitate the development of quantum software infrastructures optimized for hybrid learning and communication.

Finally, it would be worthwhile to investigate scenarios where the teacher model is large but noisy, while the student model is smaller and well-calibrated, to refine KD applications in realistic quantum environments and advance its practical utility. It would be valuable to investigate whether insights derived from large but noisy quantum systems can improve QNN implementations on smaller, more manageable devices.

\section*{Data availability}
The data supporting the findings of this study are available from the first author upon reasonable request. The theoretical results of the manuscript are reproducible from the analytical formulas and derivations presented therein. A demonstration code repository is available \footnote{\url{https://github.com/royess/QNNKDMindSpore}}. Additional code is available from the first author upon reasonable request.

\begin{acknowledgments}
This work is sponsored by CPS-Yangtze Delta Region Industrial Innovation Center of Quantum and Information Technology-MindSpore Quantum Open Fund.
Y. Y. acknowledges the National Natural Science Foundation of China Grant No.~12174216, and the Innovation Program for Quantum Science and Technology Grant No.~2021ZD0300804. Q. Z. and X. Z. acknowledges funding from Innovation Program for Quantum Science and Technology via Project 2024ZD0301900, National Natural Science Foundation of China (NSFC) via Project No. 12347104 and No. 12305030, Guangdong Basic and Applied Basic Research Foundation via Project 2023A1515012185, Hong Kong Research Grant Council (RGC) via No. 27300823, N\_HKU718/23, and R6010-23, Guangdong Provincial Quantum Science Strategic Initiative No. GDZX2303007, HKU Seed Fund for Basic Research for New Staff via Project 2201100596.
\end{acknowledgments}

\section*{Author contributions}
Y.Y., S.Q., and X.Z. initiated this work. Y.Y. and S.Q. studied the task of classifying MNIST images. Y.Y., Q.Z., and X.Z. studied the specific Hamiltonian models. Y.Y. and S.Q. carried out numerical experiments. All the authors contributed to the study of the theoretical framework and writing the manuscript.

\bibliography{qnnkd}



\end{document}